\documentstyle[12pt,aaspp4,epsf]{article}
\begin{document}
\title{The ROSAT Brightest Cluster Sample (BCS): The cluster X-ray 
		luminosity function within $z=0.3$}
\author{H.\ Ebeling\altaffilmark{1}, A.C.\ Edge, A.C.\ Fabian, S.W.\ Allen, 
        C.S.\ Crawford}
\affil{Institute of Astronomy, Madingley Road, Cambridge CB3\,0HA, UK}
\and
\author{H.\ B\"ohringer}
\affil{MPI f\"ur extraterrestrische Physik, 
             Giessenbachstr., D-85740 Garching, Germany}
\altaffiltext{1}{also: Institute for Astronomy, 2680 Woodlawn Dr, 
             Honolulu HI 96822, USA; email: {\em ebeling{@}ifa.hawaii.edu}}

\slugcomment{accepted for publication in ApJL}

\begin{abstract}
We present and discuss the X-ray luminosity function (XLF) of the
ROSAT Brightest Cluster sample (BCS), an X-ray flux limited sample of
clusters of galaxies in the northern hemisphere compiled from ROSAT
All-Sky Survey data. The BCS allows the local cluster XLF to be
determined with unprecedented accuracy over almost three decades in
X-ray luminosity and provides an important reference for searches for
cluster evolution at higher redshifts.

We find the significance of evolution in both the XLF amplitude and in
the characteristic cluster luminosity $L_X^{\star}$ to be less than
$1.8\sigma$ within the redshift range covered by our sample thereby
disproving previous claims of strong evolution within $z\la 0.2$.

\end{abstract}


\section{Introduction} 

The ROSAT Brightest Cluster sample (BCS, Ebeling et al.\markcite{bcsi}
1996b, hereafter Paper I) is a 90\% complete, flux limited
sample of the 199 X-ray brightest clusters of galaxies in the northern
hemisphere ($\delta \geq 0^{\circ}$), at high Galactic latitudes ($|b|
\geq 20^{\circ}$, with redshifts $z \leq 0.3$, fluxes higher than
$4.45\times 10^{-12}$ erg cm$^{-2}$ s$^{-1}$ and luminosities higher
than $5\times 10^{42}$ erg s$^{-1}$ in the 0.1--2.4 keV band. Second
in size only to the XBACs sample of Ebeling et al.\markcite{xbacs}
(1996a), the BCS is one of the largest statistical cluster samples
compiled at X-ray wavelengths to date. It is the only large-scale
sample available today that is not only X-ray {\em flux limited}\/ but
also X-ray {\em selected} in the sense that the BCS, unlike the XBACs,
is not limited to systems initially found in optical surveys but
contains clusters selected by their X-ray properties only.

The BCS thus represents an ideal sample for studies of the formation,
distribution and evolution of structure on the largest metric and mass
scales. Providing important constraints on the cosmological parameters
governing cluster evolution, the X-ray luminosity function (XLF) of
clusters of galaxies represents a particularly vital statistic in this
context.

Several previous studies based on much smaller samples of typically 50
clusters found the evolution in the cluster X-ray luminosity function
to be `negative' in the sense that X-ray luminous clusters are more
numerous now than they were in the
past\markcite{edge,gioia,henry,david} (Edge et al.\ 1990; Gioia et
al.\ 1990; Henry et al.\ 1992; David et al.\ 1993).  They were,
however, not only in conflict with other studies, which found no
evidence for cluster evolution (e.g.,\markcite{kowalski} Kowalski et
al.\ 1984), but also somewhat inconsistent among themselves.  The
strong evolution seen by Edge et al.\ (1990) in their sample of 46
X-ray bright clusters at high Galactic latitude and $z \leq 0.18$ is
not present in the first two redshift bins (44 clusters at $0.14\leq z
\leq 0.3$) of the sample of Gioia et al.\markcite{gioia} (1990) and
Henry et al.\markcite{henry} (1992) who find significant evolution
only at $z>0.3$. More recently, two studies found no sign of evolution
at all in the XLF of samples of Abell and ACO clusters at $z\le 0.36$
\markcite{briel} (Briel \& Henry 1993) and $z\le 0.15$ \markcite{burg}
(Burg et al.\ 1994), respectively.  However, these samples were
neither X-ray selected nor X-ray flux limited and may thus not be fair
representations of the cluster population in general. At considerably
higher redshifts ($z>0.5$) on the other hand, studies based on yet
smaller samples observed in deep X-ray pointings
\markcite{bower,castander1,castander2} (Bower et al.\ 1994; Castander
et al.\ 1994; Castander et al.\ 1995) suggest a significant drop in
the cluster space density as compared to the value observed locally.

Although the overall evidence is thus in favor of negative evolution
of the cluster XLF at least for X-ray luminous clusters at redshifts
well in excess of 0.3, the overall picture is anything but clear.
With the completion of the BCS we are now able to provide a definitive
answer to the question of whether cluster evolution is significant at
low to intermediate redshifts and, in any case, provide an accurate
determination of the local cluster XLF as a much-needed reference for
ongoing and future evolutionary studies at higher redshifts. The
implications of our findings for cosmological models of cluster
evolution will be addressed in a forthcoming paper (Ebeling et al., in
preparation). We assume $H_0=50$ km s$^{-1}$ Mpc$^{-1}$ and $q_0=0.5$
throughout this paper.

\section{The BCS XLF and its parametrization}

The BCS as published in Paper I is only 90\% complete and
corrections for incompleteness need to be applied to account for
clusters missing from the sample. In doing this we use a selection
function based on the $z, L_{\rm X}$ distribution of the serendipitous
detections in our sample (see \S 7.1 of Paper I).

We use the usual definition of the unbinned, differential XLF for a
sample with flux limit $f_{X,{\rm lim}}$:
\begin{displaymath}
    {\rm XLF}(L_X,z,f_{X,{\rm lim}}) = \frac{dn(L_X,z,f_{X,{\rm lim}})}{dL_X}
\end{displaymath}
where $dn(L_X,z,f_{X,{\rm lim}})$ is the space density of clusters
with X-ray fluxes above the flux limit and X-ray luminosities within
an interval $dL_X$ around $L_X$. Since we use an unbinned
representation, $dn$ is given for each cluster by $1/V(L_X,z,f_{X,{\rm
lim}})$, i.e., the inverse search volume defined by the luminosity
distance at which the X-ray flux from a cluster with intrinsic
luminosity $L_X$ would equal the flux limit of the sample. The
systems' X-ray temperatures as listed in Paper I are used in the
computation of K corrections.

We use a Schechter function\markcite{schechter} (Schechter 1976) of
the form
\begin{displaymath}
    \frac{dn}{dL_X} (L_X) = A \, \exp (-L_X/L_X^{\star}) \, L_X^{-\alpha}
\end{displaymath}
to model the XLF and determine $A$, $L_X^{\star}$ and $\alpha$ in a
maximum-likelihood fit to the unbinned data. 

Table~\ref{partab} gives an overview of the fit results obtained in
all five standard energy bands currently in use within the scientific
community: $0.1-2.4$, $0.5-2.0$, $0.3-3.5$, and $2-10$ keV, as well as
the pseudo-bolometric band from 0.01 to 100 keV. We discuss the results
in detail in the following paragraphs.

\begin{figure}
\epsfxsize=\textwidth
\epsffile{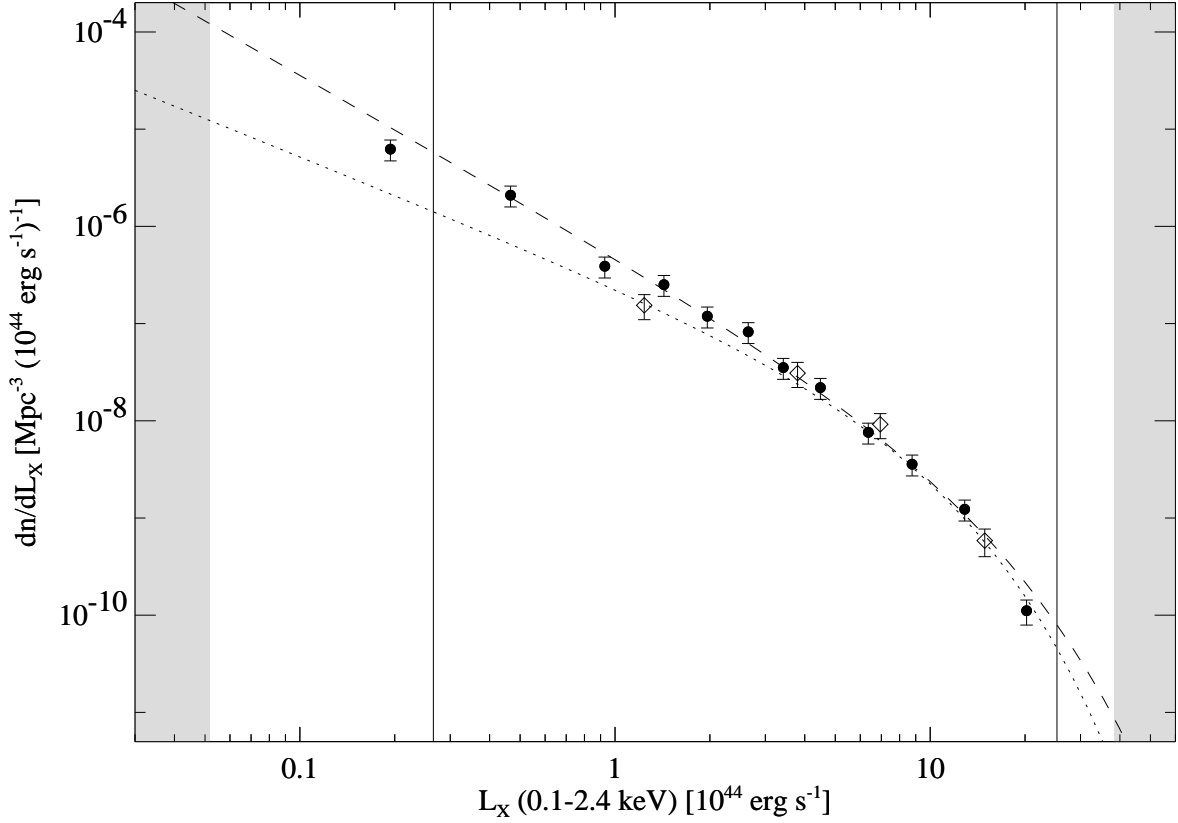}
\figcaption{The X-ray luminosity function (filled
				  circles) for the ROSAT Brightest
				  Cluster Sample in the $0.1-2.4$ keV
				  band. The data are grouped such that
				  each bin contains 17 clusters,
				  except the highest-luminosity one
				  which contains 12 clusters. The
				  dashed line shows the best Schechter
				  function fit. The shaded region
				  marks the luminosity range not
				  covered by the BCS. Overlaid is the
				  XLF of the B50 sample of Edge et
				  al.\ (1990) when converted to the
				  $0.1-2.4$ keV band and grouped such
				  that each bin contains 12 clusters,
				  except the highest-luminosity one
				  which contains 10 clusters (open
				  diamonds). The luminosity range
				  covered by the B50 in the $0.1-2.4$
				  keV band is indicated by the solid
				  vertical lines; the best Schechter
				  function fit to the unbinned B50
				  data is represented by the dotted
				  line.  \label{xlf0.1-2.4} }
\end{figure}

Figure~\ref{xlf0.1-2.4} shows the differential XLF for the BCS in the
generic $0.1-2.4$ keV band of the ROSAT observatory. We test the
robustness of the fit by comparing it to the XLF obtained for the
larger, 80\% complete BCS (unpublished, cf.~Paper I) and find
excellent agreement. 
Also shown in Figure~\ref{xlf0.1-2.4} are the XLF data points for the
high-galactic-latitude sample of Edge et al.\ (1990, hereafter B50)
(open diamonds) which is X-ray flux limited in the $2-10$ keV
band. Since all 46 B50 clusters have measured X-ray temperatures, the
conversion of their luminosities to the $0.1-2.4$ keV band of the BCS
is less inaccurate than the opposite operation, i.e., the conversion
of the BCS luminosities into the $2-10$ keV band, which relies heavily
on temperature estimates rather than measured values.  To assess the
impact of band conversion effects we make the comparison between the
XLFs of the B50 and the BCS in either band.  In the $0.1-2.4$ keV band
we find the XLF for the B50 to be, in general, in good agreement with
the BCS XLF. At low X-ray luminosities, however, the best Schechter
function fit for the B50 (the dotted line in Figure~\ref{xlf0.1-2.4}
and obtained with the same ML algorithm used throughout for the
fitting of the BCS XLFs) severely underpredicts the observed volume
density of clusters, indicating incompleteness of the B50 at $L_X \la
1\times 10^{44}$ erg s$^{-1}$ ($0.1-2.4$ keV). This is, however, not
surprising given that the B50 is, by design, only flux-limited in the
$2-10$ keV band.

\begin{figure}
\epsfxsize=\textwidth
\epsffile{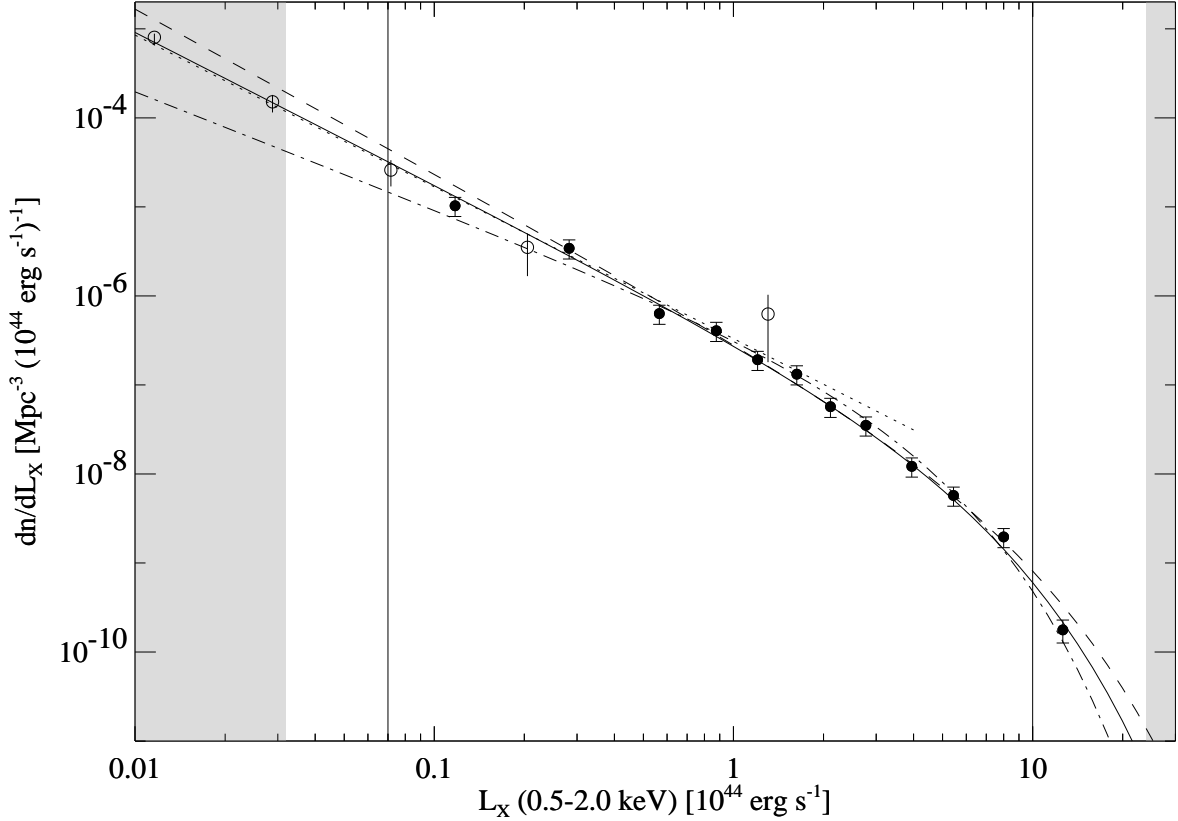}
\figcaption{The cluster X-ray luminosity
				  function in the $0.5-2.0$ keV
				  band. The dashed curve represents
				  the best Schechter function fit to
				  the BCS data (filled circles;
				  binning as in
				  Figure~\protect\ref{xlf0.1-2.4}).
				  The open circles and the dotted line
				  show the group XLF and the
				  corresponding best-fitting power law
				  from BLL. The solid line represents
				  the best Schechter function fit to
				  the XLF obtained by combining the
				  binned BCS data with the first 4
				  data points of the groups XLF as
				  determined by BLL. The dot-dashed
				  line, finally, shows the XLF
				  determined by De Grandi (1996) for a
				  sample of 111 clusters selected from
				  ROSAT data in the southern
				  hemisphere. The luminosity range of
				  this sample is indicated by the
				  solid vertical lines (De Grandi, private
				  communication).
				  \label{xlf0.5-2.0} }
\end{figure}

Figure~\ref{xlf0.5-2.0} shows the BCS XLF in the ROSAT hard band
covering the energy range from 0.5 to 2.0 keV. Note that, like for all
other energy bands discussed in the following, the conversion from the
original BCS band ($0.1 - 2.4$ keV) to the ROSAT hard band was
performed for each cluster individually using the X-ray temperatures
and Galactic column densities given in Table~2 of Paper I and assuming
a metallicity of 0.3 of the solar value.  Also shown in
Figure~\ref{xlf0.5-2.0} are the data of the XLF of groups and poor
clusters of galaxies as presented by Burns et al.\markcite{burns}
(1996, hereafter BLL) and their best power-law XLF. Note the good
agreement between the two samples as well as the respective XLF models
in the overlap region between $3\times 10^{42}$ erg s$^{-1}$ (the
lowest luminosity of any BCS cluster) and $2.6\times 10^{43}$ erg
s$^{-1}$ (the highest luminosity of BLL's poor cluster
sample)\footnote{We have tested the compatibility of the two samples
by comparing the luminosities of the four clusters contained in both
the BCS and the BLL sample. We find agreement to within 5\% between
the respective luminosities for all but one cluster (MKW8) for which
the BLL luminosity falls short of the BCS value by 40\%.}.  Finally,
we overlay in Fig.~\ref{xlf0.5-2.0} the best fitting Schechter
function XLF determined by De Grandi (1996) for the 111 clusters of
the BSGC-KP sample, a subset of a larger ROSAT cluster sample under
compilation in the southern hemisphere (Guzzo et al.\ 1995).  For
luminosities in excess of $\sim 2\times 10^{43}$ erg s$^{-1}$ De
Grandi's XLF is in very good agreement with the best-fitting BCS
Schechter function. Below this value the BSGC-KP XLF falls
increasingly below both the BLL and the BCS fits, reaching a deviation
of a factor of 2 and 3, respectively, at the lower end of the BSGC-KP
luminosity range.

\begin{figure}
\epsfxsize=\textwidth
\epsffile{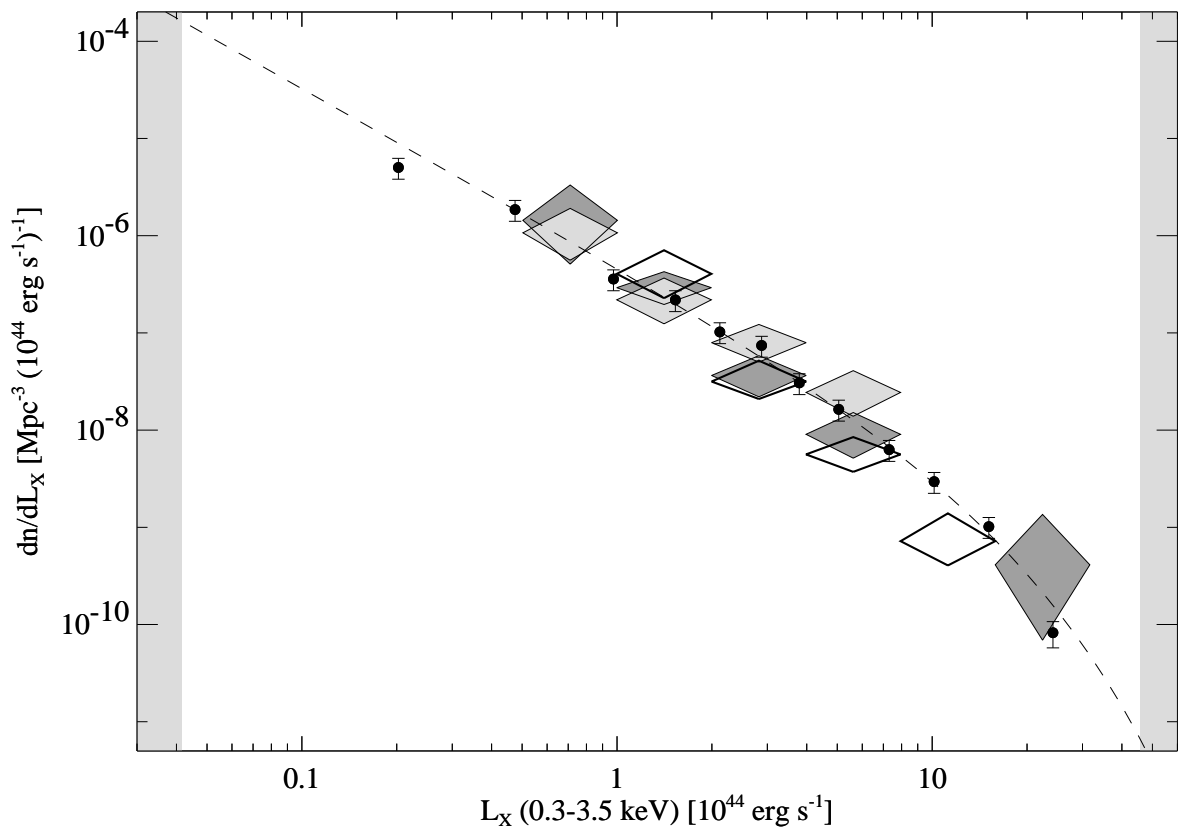}
\figcaption{The cluster X-ray luminosity
				  function in the $0.3-3.5$ keV
				  band. The dashed curve represents
				  the best Schechter function fit to
				  the BCS data (filled circles;
				  binning as in
				  Figure~\protect\ref{xlf0.1-2.4}).
				  The diamonds in light and dark
				  shading show the EMSS XLF of Henry
				  et al.\ (1992) in the $0.14\le z <
				  0.2$ and the $0.2\le z <0.3$
				  redshift shell, respectively, while
				  the EMSS XLF in the $0.3\le z < 0.6$
				  shell is represented by open
				  diamonds.  \label{xlf0.3-3.5} }

\end{figure}

The reference in the $0.3-3.5$ keV band is the EMSS cluster sample of
Henry et al.\ (1992). Figure~\ref{xlf0.3-3.5} shows the BCS XLF in the
$0.3-3.5$ keV band with the EMSS data points from the first two
redshift shells ($0.14\le z <0.2$ and $0.2 \le z <0.3$) overlaid as
shaded diamonds. (We remind the reader that the power law descriptions
of the EMSS XLF found by Henry and coworkers in these shells are
consistent with one another within the errors, so that the comparison
with the whole of the BCS made in Figure~\ref{xlf0.3-3.5} is
legitimate.)  Note the very good agreement between the EMSS and BCS
luminosity functions, but also the much higher accuracy provided by
the BCS.  Also shown in Figure~\ref{xlf0.3-3.5} is the XLF from the
third EMSS redshift shell ($0.3 \le z < 0.6$). Contrary to our
findings for the EMSS XLF for clusters at $0.14 \le z < 0.3$, the EMSS
XLF of these high-redshift systems differs noticeably from the BCS XLF
as the local reference.  A detailed re-assessment of the significance
of the evolution implied by this discrepancy is presented by Ebeling
et al.\ (in preparation).

\begin{figure}
\epsfxsize=\textwidth
\epsffile{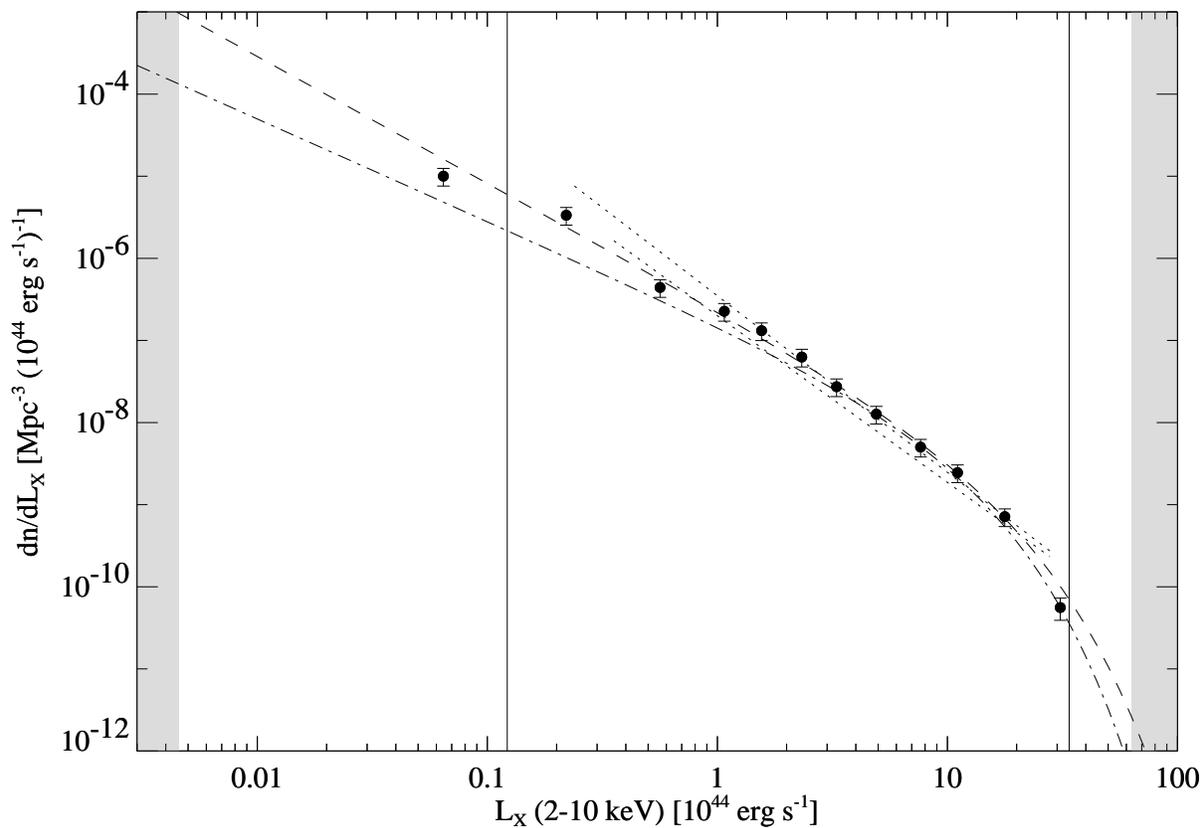}
\figcaption{The cluster X-ray luminosity function in
				  the $2-10$ keV band. The dashed
				  curve represents the best
				  Schechter function fit to the BCS data
				  (filled circles; binning as in
				  Figure~\protect\ref{xlf0.1-2.4}).
				  The two dotted lines show the
				  power-law representations of the XLF
				  of Piccinotti et al.\ (1982) with
				  and without the Virgo cluster,
				  respectively. The dot-dashed line,
				  finally, represents our best
				  Schechter function fit to the B50
				  data of Edge et al.\ (1990) which
				  covers the luminosity range bounded
				  by the solid vertical lines.
				  \label{xlf2-10} }

\end{figure}

Although, in Table~\ref{partab}, we do list the results of a Schechter
function fit to the BCS data in the $2-10$ keV band, the BCS XLF in
this hard energy band should be regarded with caution. The $2-10$ keV
band has hardly any overlap with the $0.1-2.4$ keV BCS detection band,
which renders a flux conversion that is largely based on estimated
temperatures a dangerous enterprise. Keeping this caveat in mind we
nonetheless find the power law description from Piccinotti et al.\
(1982) to be in good agreement with the BCS XLF irrespective of
whether the Virgo cluster is included or not (Figure~\ref{xlf2-10}).

The best fit parameters from the Schechter function fit to the B50
data published by Edge et al.\ (1990), on the other hand, provide an
unacceptable fit to the BCS XLF at luminosities in excess of about
$1\times 10^{45}$ erg s$^{-1}$, where the fit given by Edge et al.\
falls significantly below the BCS data. This failure of the original
B50 Schechter function fit to describe the BCS XLF is due to an error in the
volume calculation in the work of Edge and coworkers. Fitting the B50
data with our maximum likelihood algorithm we find
$A=1.59^{+0.38}_{-0.33}\times 10^{-7}$ Mpc$^{-3}\,(10^{44}$ erg
s$^{-1})^{\alpha-1}$, $L_X^{\star}=8.46^{+2.69}_{-1.83}\times 10^{44}$
erg s$^{-1}$, and $\alpha=1.25^{+0.19}_{-0.20}$, in good agreement
with the results for the BCS (see Figure~\ref{xlf2-10} and
Table~\ref{partab}).

\section{Evidence of evolution in the BCS XLF}

We search for evidence of evolution by splitting the BCS at a redshift
$z_{\rm sep}$ and independently fitting Schechter functions to the
data in the two redshift shells thus created. Care has to be taken not
to na\"{\i}vely misinterpret every statistically significant
difference between the best fit parameters in the two shells as
signature of evolution. In order to avoid effects due to large scale
structure, we vary $z_{\rm sep}$ from 0.05 to 0.2 and look for trends
that are robust over a range of $z_{\rm sep}$ values. Since, due to
the flux-limited nature of our sample, the low-luminosity end of the
XLF, and thus $\alpha$, is ill-constrained for the high-redshift
subsample once $z_{\rm sep}$ exceeds $z\sim 0.1$, we fix the power law
slope $\alpha$ at its overall best-fit value of 1.85 in the
maximum-likelihood fits to the data of either subsample.  With
$\alpha$ frozen we are thus left with two free parameters, the
normalization $A$ and the characteristic luminosity $L_X^{\star}$.

\begin{figure}
\epsfxsize=0.7\textwidth
\hspace*{2cm}\epsffile{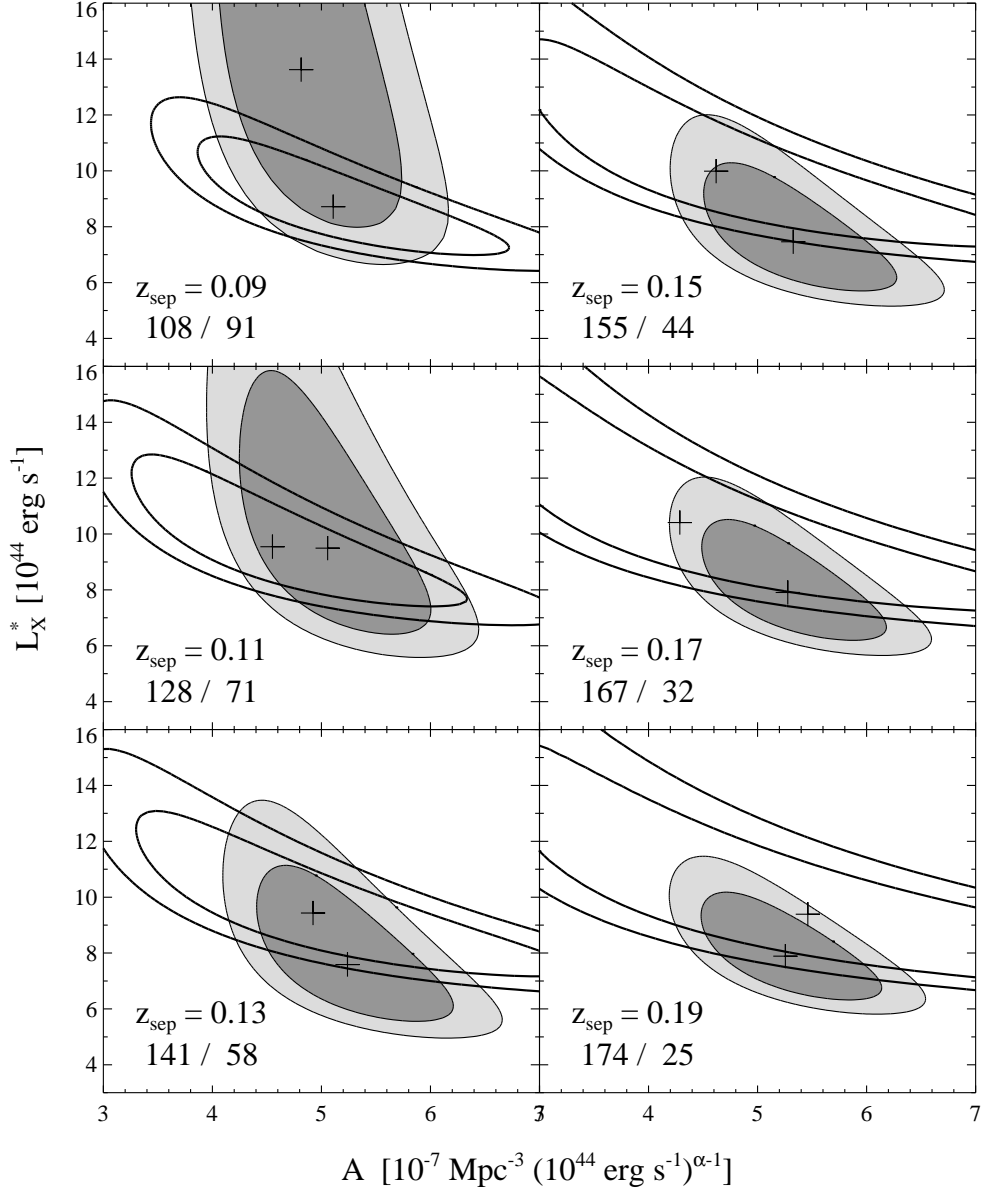}
\vskip2cm
\figcaption{Likelihood ratio contours
				  (68 and 90\% confidence for two 
				  interesting parameters, i.e., $\Delta
				  C=2.30,4.61$) in $A$ and
   				  $L_X^{\star}$ for BCS subsamples
 				  obtained by splitting the sample
 				  at an intermediate redshift
				  $z_{\rm sep}$. The various values
				  of $z_{\rm sep}$ are shown in the
				  lower left corner of each plot as
				  are the numbers of clusters in
				  the low- and the high-redshift
				  subsample, respectively.  Filled
				  contours correspond to the
				  low-redshift subsamples; the
				  contours for the high-redshift
				  subsamples are shown as bold
				  solid lines.  In the Schechter
				  function fits, $\alpha$ was kept
				  frozen at its global best-fit
				  value of $1.85$.
				  \label{evol1} }
\end{figure}

Figure~\ref{evol1} shows the contours of the C statistic (which is
$\chi^2$ distributed) of $A$ and $L_X^{\star}$ for some of these
low-redshift and high-redshift subsamples of the BCS as a function of
$z_{\rm sep}$. While differences found at $z_{\rm sep}\la 0.1$ can be
attributed entirely to large-scale structure, a significant decrease
in $A$ or $L_X^{\star}$ at higher values of $z_{\rm sep}$ would be
indicative of negative evolution. As, for $0.05\leq z_{\rm sep}\leq
0.2$, the 68\% confidence contours of the low-redshift and
high-redshift subsamples overlap, we conclude that there is no
significant evolution in either the amplitude $A$ of the cluster XLF
or the characteristic luminosity $L_X^{\star}$ for values of $z_{\rm
sep}$ up to 0.2. Since, in the high-redshift subsamples with $z_{\rm
sep} \ga 0.16$, $A$ becomes ill-constrained and increasingly strongly
correlated with $L_X^{\star}$, we also tested for evolution only in
$L_X^{\star}$ by holding $A$ constant at its overall best-fit value of
$4.33 \times 10^{-7}$ Mpc$^{-3}\,(10^{44}$ erg s$^{-1})^{\alpha-1}$, a
value well within the 68\% confidence contours of all fits shown in
Fig.~\ref{evol1}. We find the variations in $L_X^{\star}$ to be
smaller than 30(37)\% for $0.1 < z_{\rm sep} \le 0.20(0.22)$
which is less than $1.6(1.8)\sigma$ significant, confirming the
no-evolution result of Fig.~\ref{evol1}. 

As an independent check, we also looked for variations in $V/V_{\rm
max}$ as a function of both $z$ and $L_X$ --- and found none. A
Kolmogorov-Smirnov test finds the distribution of $V/V_{\rm max}$
values (whose median is 0.47) to be consistent with uniformity at the
greater than 74\% confidence level suggesting again that the
cluster space density of the BCS is homogeneous out to the limiting
redshift of $z=0.3$.

\section{Conclusions}

Using the ROSAT Brightest Cluster Sample (BCS) as presented by Ebeling
et al.\ (1996b) we have established the local X-ray luminosity
function (XLF) of clusters of galaxies within $z=0.3$ with
unprecedented accuracy. We find the XLF to be well described by a
Schechter function whose free parameters $A$, $L_X^{\star}$, and
$\alpha$ we determine in a maximum-likelihood fit for all X-ray energy
bands currently used within the community. Comparing our results with
previous measurements of the cluster XLF we find very good agreement
with the work of Piccinotti et al.\ (1982), Henry et al.\ (1992), and
Burns et al.\ (1996), as well as with the XLF for the B50 sample of
Edge et al.\ (1990) when the same maximum likelihood algorithm is used
to determine the best Schechter function fit.

We find no significant variations in the amplitude or the characteristic
luminosity of the best-fitting Schechter function as a function of
redshift. Also, the distribution of $V/V_{\rm max}$ values is consistent
at the 74\% confidence level with a non-evolving space density
of clusters out to $z=0.3$. Our findings do thus not confirm the claim
of strong evolution at $z\la 0.2$ made by Edge and coworkers but
support the notion of Ebeling et al.\ (1995) that the apparent
signature of evolution in the B50 sample is due to a combination of
its high X-ray flux limit in the $2-10$ keV band and a pronounced, if
statistically insignificant, dearth of very X-ray luminous clusters
around a redshift of about 0.15.

\acknowledgements H.E. thanks Pat Henry for helpful discussions about
maximum-likelihood fitting and cluster evolution.  H.E. acknowledges
financial support from a European Union EARA Fellowship and SAO
contract SV4-64008. A.C.E., A.C.F. and S.W.A. thank the Royal Society for
support. C.S.C. acknowledges financial support from a PPARC Advanced
Fellowship. H.B. thanks the BMFT for financial support through the 
Verbundforschung programme.

\newpage
\begin{deluxetable}{llll} 
\tablecolumns{4}
\tablewidth{0pc} 
\tablecaption{Best fit values of the Schechter function parameters
		$A$, $L_X^{\star}$, and $\alpha$ \label{partab}} 
\tablehead{
\colhead{Energy Band} & 
\colhead{$A^{\dagger}$} & 
\colhead{$L_X^{\star}{}^{\ddagger}$} & 
\colhead{$\alpha$}}

\startdata 
0.1--2.4 keV & 5.06$^{+0.50}_{-0.46}$ & $9.10^{+2.06}_{-1.49}$   & $1.85^{+0.09}_{-0.09}$ \\ 
0.5--2.0 keV & 3.32$^{+0.36}_{-0.33}$ & $5.70^{+1.29}_{-0.93}$   & $1.85^{+0.09}_{-0.09}$ \\  
0.3--3.5 keV & 4.95$^{+0.48}_{-0.45}$ & $\!\!10.7^{+2.4}_{-1.8}$ & $1.82^{+0.08}_{-0.09}$ \\ 
  2--10  keV & 2.35$^{+0.22}_{-0.18}$ & $\!\!12.6^{+2.2}_{-1.8}$ & $1.54^{+0.05}_{-0.06}$ \\ 
  bolometric & 6.41$^{+0.70}_{-0.61}$ & $\!\!37.2^{+16.4}_{-3.8}$& $1.84^{+0.09}_{-0.04}$ \\ 
\enddata
\tablenotetext{\dagger}
      {in units of 10$^{-7}$ Mpc$^{-3}\,(10^{44}$ erg s$^{-1})^{\alpha-1}$}
\tablenotetext{\ddagger}
      {in units of 10$^{44}$ erg s$^{-1}$}
\tablenotetext{}{Errors given correspond to 68\% confidence for one interesting parameter ($\Delta C=1$).}
\end{deluxetable}

\newpage

\end{document}